\documentclass[aps,prl,twocolumn,longbibliography]{revtex4-1}
\usepackage{graphicx}
\usepackage{natbib}
\usepackage{amsmath}

\newcommand{\TJWat}{IBM T.J. Watson Research Center, Yorktown Heights, NY 10598, USA}

\newcommand{\figfolder}[1]{}

\usepackage[usenames,dvipsnames]{xcolor}
\usepackage[dvips,letterpaper=true,pagebackref=false]{hyperref}
\hypersetup{
	pdfnewwindow=true,
	colorlinks=true,
	linkcolor=MidnightBlue,
	citecolor=Magenta,
	urlcolor=RoyalBlue
}

\begin{document}

\title{Three Qubit Randomized Benchmarking}

\author{David C. McKay}
\email{dcmckay@us.ibm.com}
\author{Sarah Sheldon}
\author{John A. Smolin}
\author{Jerry M. Chow}
\author{Jay M. Gambetta}
\affiliation{\TJWat}
\date{\today}

\begin{abstract}
	As quantum circuits increase in size, it is critical to establish scalable multiqubit fidelity metrics. Here we investigate, for the first time, three-qubit randomized benchmarking (RB) on a quantum device consisting of three fixed-frequency transmon qubits with pairwise microwave-activated interactions (cross-resonance). We measure a three-qubit error per Clifford of 0.106 for all-to-all gate connectivity and 0.207 for linear gate connectivity. Furthermore, by introducing mixed dimensionality simultaneous RB --- simultaneous one- and two-qubit RB --- we show that the three-qubit errors can be predicted from the one- and two-qubit errors. However, by introducing certain coherent errors to the gates we can increase the three-qubit error to 0.302, an increase that is not predicted by a proportionate increase in the one- and two-qubit errors from simultaneous RB. This demonstrates three-qubit RB as a unique multiqubit metric.
\end{abstract}

\maketitle

Quantum circuits are being built with an increasingly larger number of qubits and, accordingly, the problem of characterization is becoming more acute. The fundamental reason for quantum speedups --- the exponential growth of the state space with the number of qubits --- means that tomographic methods for reconstructing the system will require exponential resources. Indeed, the number of required measurements for quantum process tomography scales as $16^{n}$~\cite{chuang:1997} where $n$ is the number of qubits. To avoid scaling issues, methods have focused on characterizing the primitive set of gates used to construct the universal gateset. At minimum, for $n$ qubits, this set contains several one-qubit gates for all $n$ qubits and $n-1$ two-qubit gates~\cite{nielsen:2000}. But how good is the assumption that multiqubit algorithmic fidelities will be predicted by the fidelities of the gate primitives measured in isolation? There are strong indications that this assumption fails due to crosstalk and addressability errors. For example, to realize logical fault-tolerant qubits using surface code algorithms requires constructing local five-qubit gates via sequential application of two-qubit CNOT gates in parallel across a circuit with many qubits. Surface codes are predicted to have a high threshold for correcting errors, but they are typically simulated with correlated noise only between qubits for which there is a direct gate~\cite{fowler:2012}. In a recent five-qubit test of a logical qubit, the logical qubit fidelity was greatly improved by compensating for ZZ terms to spectator (i.e., non-participating neighboring) qubits during the two-qubit gate~\cite{takita:2017}. In addition, several studies have observed that algorithmic and primitive gate fidelity do not always agree. For example when four algorithms were run on two different five-qubit processors there was no definitive agreement from primitive to algorithmic fidelity~\cite{linke:2017}. In a five-qubit device with measured two-qubit gate fidelities of 0.99, the state fidelity of a five-qubit GHZ state was 0.82 after applying four two-qubit gates~\cite{barends:2014}. Therefore, to predict the true algorithmic fidelity we need to measure multiqubit fidelity metrics.

Fortunately, the issue of scaling can be circumvented if the goal is to characterize a process based on a few measures, e.g., average gate fidelity. Based on this idea, there have been several proposed techniques such as Monte Carlo sampling~\cite{flammia:2011,dasilva:2011}, compressed sensing~\cite{gross:2010}, matrix product state tomography~\cite{cramer:2010}, and twirling protocols~\cite{moussa:2012}. These techniques have been applied to perform tomography on a 6-qubit photonic state~\cite{schwemmer:2014},  measure the fidelity of a 7-qubit NMR gate~\cite{lu:2015}, to reconstruct a 7-qubit code state~\cite{riofrio:2017} and to characterize a 14-qubit ion trap~\cite{lanyon:2017}. However, a common drawback to these techniques is that the result is sensitive to preparation and measurement errors, sometimes exponentially so. In addition, Refs~\cite{riofrio:2017,lu:2015} characterized the state of a multiqubit system, but not the underlying gates. These problems are addressed by randomized benchmarking~\cite{knill:2008,magesan:2011} (RB), where sequences of random Clifford gates equaling the identity operator are applied to a set of qubits. Qubit polarization versus sequence length decays exponentially and the decay constant is a simple measure of the average fidelity of the Clifford set independent of preparation and measurement errors. RB is a method widely used to characterize gates in superconducting circuits~\cite{barends:2014,mckay:2017,chow:2009,corcoles:2013}, ion-traps~\cite{gaebler:2016,ballance:2016,gaebler:2012,knill:2008}, neutral-atom-traps~\cite{olmschenk:2010}, NMR systems~\cite{ryan:2009} and for solid-state spin qubits~\cite{veldhorst:2014}. Extensions to RB have been proposed and implemented to measure specific gate error via interleaving~\cite{magesan:2012}, purity~\cite{mckay:2016,wallman:2015} and leakage~\cite{wood:2017,wallman:2016}.

RB is designed to address fidelities in multiqubit systems in two ways. For one, RB over the full $n$-qubit space can be performed by constructing sequences from the $n$-qubit Clifford group. Additionally, the $n$-qubit space can be subdivided into sets of qubits $\{n_i\}$ and $n_i$-qubit RB performed in each subset simultaneously~\cite{gambetta:2012}. Both methods give metrics of fidelity in the $n$-qubit space. Despite the availability of these two methods, there has been no demonstration of RB with $n>2$ since it is viewed as sufficient to characterize only the primitive gateset. Here we show, for the first time, a variety of three-qubit RB combinations in a three-qubit fixed-frequency superconducting device. For all-to-all gate connectivity we measure a three-qubit error per Clifford (3Q EPC) of 0.106, which is well-predicted by the primitive gate errors from two-qubit/one-qubit simultaneous RB. However, we find a strong dependence on calibration procedure that is not apparent a priori. For one such calibration the error increases to 0.302. Importantly, this increase in error is not predicted by a commensurate increase in the primitive gate errors as measured from simultaneous RB. We also show the importance of connectivity in devices as the 3Q EPC increases to 0.207 when we limit the device to have linear gate connectivity.

\begin{figure}
\includegraphics[width=0.4\textwidth]{\figfolder{1}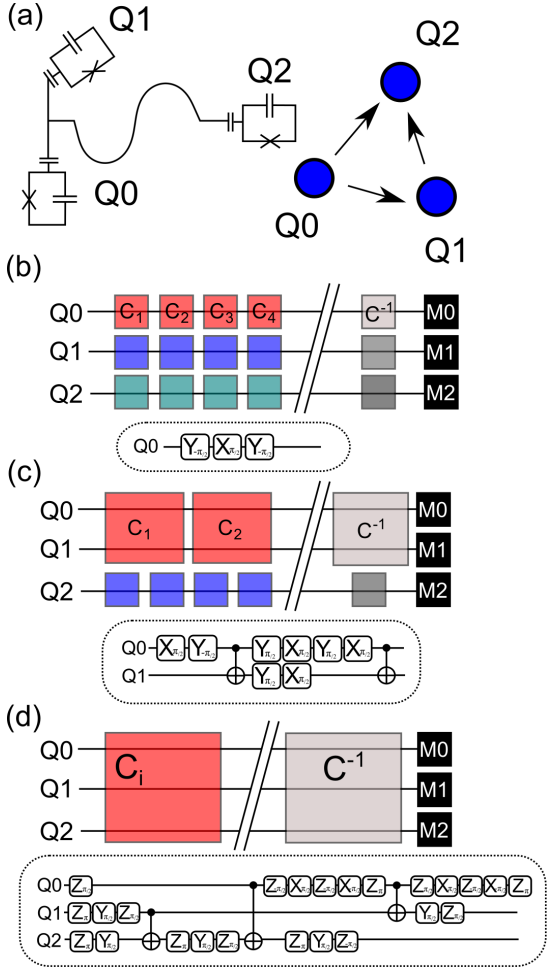}
\caption{(Color Online) (a) Schematic of the experimental setup and connectivity of the CNOT 2Q gates (control$\rightarrow$target). (b) 1Q simultaneous RB $\{[0],[1],[2]\}$, (c) 2Q-1Q simultaneous RB $\{[0,1],[2]\}$ and (c) 3Q RB $\{[0,1,2]\}$. Under each is a sample (b) 1Q (c) 2Q and (d) 3Q Clifford gate. \label{fig:1}}
\end{figure}

Before describing our experiment in detail we first provide a brief summary of the RB method; a detailed discussion of RB can be found in Ref~\cite{magesan:2012b}. The main idea is to construct an $m$-length sequence of random $n$-qubit Clifford gates $\prod_{i}^{m-1} \{C_{n,i}\} = \tilde{C}_{n,m-1}$ which is appended by the inverse of the sequence $\tilde{C}_{n,m-1}^{-1}$. Such an inverse is efficiently calculated by the Gottesman-Knill theorem~\cite{gottesman:1998}. Starting in the state $|0\rangle^{\otimes n}$ and applying the full sequence of Clifford gates, we then measure the population in $|0\rangle$ of each qubit. This procedure is repeated $l$ times for different random sequences, which in the limit of large $l$, twirls the error map to a depolarizing error map $\Lambda[\rho]=\alpha \rho + (1-\alpha) \mathcal{I}/d$ where $p=1-\alpha$ is the depolarizing probability. The population in $|0\rangle$ versus the sequence length fits to an exponential decay $A\alpha^{m}+B$ and the average error over the Clifford gates is
\begin{equation}
	EPC = \frac{2^n-1}{2^n} \left(1-\alpha\right), \label{eqn:epc}
\end{equation}
(for a wide variety of noise models~\cite{epstein:2014,wallman:2017,proctor:2017}). State preparation and measurement errors do not affect the decay constant. The number of gates in the Clifford group grows superexponentially --- there are 24 one-qubit gates, 11520 two-qubit gates and 92897280 three-qubit gates~\cite{ozols:2008}. However, the method only requires fair sampling from this set. Each gate is constructed from a set of primitive gates and the exact number of 1Q and 2Q gates required depends on the basis used. In this work, our 2Q gate is a controlled NOT (CNOT$_{ij}$) where $i$ is the control and $j$ is the target. We generate our 1Q and 2Q Clifford gates using the set of 1Q gates $\{I,X_{\pi/2},X_{-\pi/2},Y_{\pi/2},Y_{-\pi/2}\}$ where $P_{\theta}=e^{-i\theta/2 \hat{P}}$. With this gate set there are 2.2083 1Q primitive gates per 1Q Clifford and 1.5 CNOT gates and 12.2167 1Q gates per 2Q Clifford. To generate the 3Q Cliffords we use the set of 1Q gates $\{X_{\pi/2},X_{-\pi/2},Y_{-\pi/2}\}$ plus arbitrary Z rotations, which are software defined~\cite{mckay:2017}; this is the set used by the Qiskit compiler~\cite{qiskit}. For all-to-all connectivity there are 3.5 CNOT gates and 11.6 1Q gates (counting only $X$ and $Y$). We use the Qiskit compiler to change the connectivity by removing one of the CNOT gates which results in an average of 7.7 CNOT gates and 18.4 1Q gates per 3Q Clifford. Sample 1Q, 2Q and 3Q Cliffords are shown in Fig.~\ref{fig:1}.

In the case of multiqubit systems, RB may be performed on the full $n$-qubits (as detailed above), or on subsets of the system. For example, it is common to perform 2Q RB on the subset of two-qubits defining a CNOT gate while the other qubits are quiescent. As explained in Ref~\cite{gambetta:2012}, this RB data will not necessarily decay exponentially because the other qubit subspaces are not twirled. Subsets are more rigorously characterized by simultaneous RB, which also measures some level of crosstalk error since all qubits are active. Herein we will use the notation $\{[i,j],...,[k]\}$ to denote benchmarking where the $m^{\mathrm{th}}$ set of $n_m$ qubits is performing independent $n_m$-qubit RB. For example $\{[0],[1,2]\}$ would indicate 1Q RB on qubit 0 and 2Q RB on qubits 1 and 2. The different combinations for three-qubits are shown in Fig.~\ref{fig:1}.

To test 3Q RB we use a device comprised of three fixed-frequency superconducting transmon qubits (Q0,Q1,Q2) of frequencies (5.353,5.291,5.237)~GHz coupled to a common 6.17GHz bus resonator. Our 1Q gates are 44.8~ns wide DRAG shaped microwave pulses~\cite{motzoi:2009}. Our 2Q gates are Gaussian smoothed square microwave pulses applied to a qubit (the control) at the frequency of one of the other qubits (the target). This activates a cross-resonance interaction, which can be tuned to build a composite pulse CNOT gate of 240~ns; details are found in Ref~\cite{sheldon:2016}. A schematic of the device and CNOT connectivity is shown in Fig.~\ref{fig:1}. More device details are given in Ref.~\cite{mckay:2018}.

\begin{figure}
\includegraphics[width=0.4\textwidth]{\figfolder{2}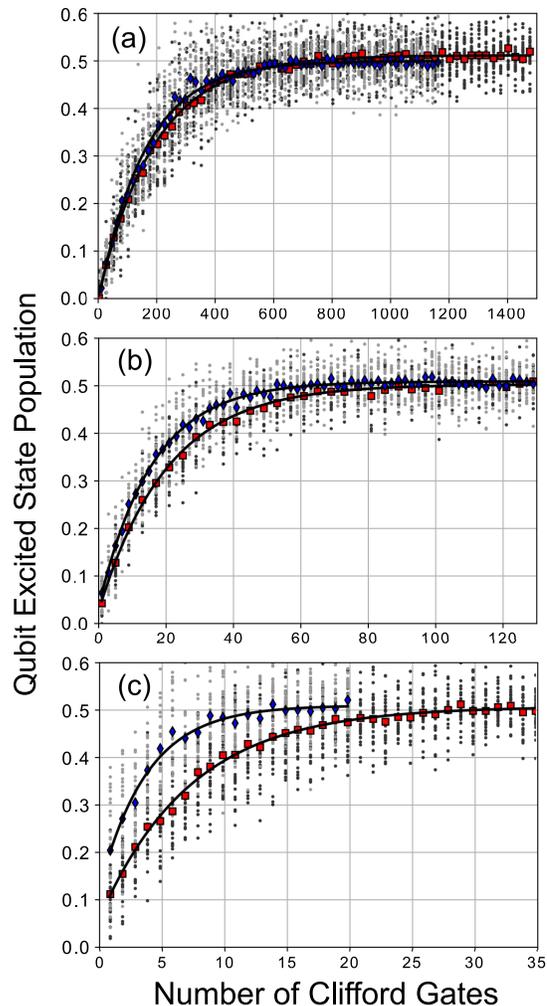}
\caption{(Color Online) Qubit 0 experimental data from different RB sequences for calibration A. Black lines are exponential fits to the data and the gray points are from the individual seeds. Red squares (blue diamonds) are the averages over these seeds for the light gray (dark gray) points. (a) 1Q RB from simultaneous 1Q (red squares) and 2Q-1Q RB (blue diamonds). (b) 2Q RB for the 01 pair performed in isolation (red square) and simultaneously with 1Q RB on Q2 (blue diamonds). (c) 3Q RB for all-to-all connectivity (red squares) and for limited (no CNOT$_{12}$) connectivity (blue squares). The decay parameters from these fits are given in Ref.~\cite{mckay:2018}.  \label{fig:2}}
\end{figure}

For our three-qubit system we consider 8 possible RB combinations: simultaneous 1Q RB ($\{[0],[1],[2]\}$), separate 2Q RB ($\{[0,1]\}$,$\{[0,2]\}$,$\{[1,2]\}$), simultaneous 2Q RB and 1Q RB (2Q-1Q RB) ($\{[0,1],[2]\}$,$\{[0,2],[1]\}$,$\{[1,2],[0]\}$) and, finally, 3Q RB ($\{[0,1,2]\}$). For each combination we perform $l=30$ averages (except for separate 2Q RB where $l=20$).  For simultaneous RB we attempt to match the sequence lengths on the different subsystems. This occurs naturally (on average) for simultaneous 1Q RB because all the 1Q gates are the same width. For 2Q-1Q simultaneous RB we use a fixed ratio of 9 between the number of 1Q Clifford gates and 2Q Clifford gates. As previously discussed we measure 3Q RB with all-to-all and limited gate connectivity. We perform these RB sequences under two different calibration procedures. In procedure A we calibrate the 1Q gate parameters simultaneously, e.g., qubit frequency, pulse amplitude and drag amplitude. In procedure B we calibrate the 1Q gate parameters individually. In both cases we calibrate the 2Q gates separately. To give a sense of the types of curves produced from 1Q, 2Q and 3Q RB, a subset of the data from calibration A is shown in Fig.~\ref{fig:2}. The errors from the full RB set and for both calibrations are summarized in Table~\ref{tab:meas}.

\begin{table*}
\begin{tabular}{p{0.5\textwidth}|c|c|}  & Cal. A & Cal. B \\ \hline \hline
$T_1$ & [29,50,39]~$\mu$s & [42,47,35]~$\mu$s \\ \hline
$T_2$ & [39,75,59]~$\mu$s & [61,74,46]~$\mu$s \\ \hline
1Q EPG Coherence Limit & [6.5,3.5,4.4]$\times 10^{-4}$ & [4.2,3.6,5.4]$\times 10^{-4}$ \\ \hline
1Q EPG from $\{[0],[1],[2]\}$ RB & [1.12(2),0.86(1),1.22(2)]$\times 10^{-3}$ & [1.40(5),0.81(1),1.66(4)]$\times 10^{-3}$ \\ \hline
1Q EPG from $\{[i],[j,k]\}$ RB & [1.41(3),0.95(2),1.35(2)]$\times 10^{-3}$ & [1.68(4),0.95(2),1.54(3)]$\times 10^{-3}$ \\ \hline
2Q EPG Coherence Limit & [6,7,5]$\times 10^{-3}$ & [5,6,6]$\times 10^{-3}$ \\ \hline
2Q EPG from $\{[i,j]\}$ RB  & [1.26(7),1.15(8),2.8(2)]$\times 10^{-2}$ & [0.86(5),2.8(1),0.92(7)]$\times 10^{-2}$ \\ \hline
2Q EPG from $\{[i,j],[k]\}$ RB & [1.89(6),1.62(6),1.74(7)]$\times 10^{-2}$ & [2.45(8),4.2(2),4.3(2)]$\times 10^{-2}$ \\ \hline
3Q EPC from $\{[0,1,2]\}$ RB (all-to-all) & 0.106(2) & 0.302(6) \\ \hline
3Q EPC from $\{[0,1,2]\}$ RB (omit CNOT$_{12}$) & 0.207(3) & N/A \\ \hline
\end{tabular}
\caption{EPG (error per gate) and EPC (error per Clifford) from different RB experiments in [Q0,Q1,Q2] order for 1Q (one-qubit) EPG and in order [CNOT$_{01}$, CNOT$_{02}$, CNOT$_{12}$] for the 2Q (two-qubit) EPG. 1Q EPG is the error per gate averaged over the set indicated in the main text. 2Q EPG is calculated from the 2Q EPC assuming the 1Q EPG from $\{[0],[1],[2]\}$ benchmarking (see~\cite{mckay:2018} for details of this calculation). 3Q EPC omitting CNOT$_{12}$ for calibration B is N/A because the error was too high to properly fit the data. The coherence limited errors are calculated assuming only errors from $T_1$ and $T_2$. Variability in $T_1$ and $T_2$ between the calibrations is due to drift over the approximately three days between experiments. Errors reflect the uncertainty in the fit parameters.  \label{tab:meas}}
\end{table*}

\begin{table}
\begin{tabular}{p{0.15\textwidth}|p{0.1\textwidth}|p{0.1\textwidth}|p{0.1\textwidth}|}  & \multicolumn{2}{c|}{Cal. A} & Cal. B \\ \hline
 & All-to-All & Omit CNOT$_{12}$ & All-to-All \\ \hline \hline
	3Q EPC from RB & \bf{0.106(2)} & \bf{0.207(3)} & \bf{0.302(6)} \\ \hline \hline
Coherence Limit & 0.044 & 0.094 & 0.041 \\ \hline
	3Q EPC Predicted from $\{[i],[j,k]\}$ RB & 0.115(4) & 0.226(6) & 0.187(7) \\ \hline
\end{tabular}
\caption{Predicted 3Q EPC from 1Q and 2Q EPG numbers listed in Table~\ref{tab:meas} by applying Eqn.~\ref{eqn:3qepc}. See the main text for a detailed dicussion of the calculation. \label{tab:pred}}
\end{table}

The data from Table~\ref{tab:meas} clearly demonstrate the difference between benchmarking isolated 2Q gates versus $\{[i],[j,k]\}$ RB (2Q-1Q simultaneous RB). Almost all errors from 2Q-1Q RB are worse which is consistent with increased crosstalk. There is one exception, CNOT$_{12}$, for calibration A which decreases from 2.8$\times10^{-2}$ to 1.74$\times10^{-2}$. This highlights the difference between the calibration procedures, mainly that they result in different calibrated values for the qubit frequency. The qubit frequencies in calibration A are shifted by the average ZZ interaction between pairs (ZZ$_{01}$=20~kHz, ZZ$_{02}$=352~kHz and ZZ$_{12}$=114~kHz). Since the ZZ$_{02}$ shift is calibrated into the frequency of Q2 for calibration A, there is a Z error when benchmarking CNOT$_{12}$ if Q0 is in the ground state; the opposite is true for calibration B and so the standalone CNOT$_{12}$ RB error is very low (0.92$\times10^{-2}$). Although there is only a subtle difference between the calibration procedures, there is a large difference between the 3Q RB errors illustrating how 3Q RB can be a sensitive probe of such calibration procedures on algorithmic fidelity. Overall, calibrating the average ZZ into the qubit frequencies maximizes 3Q fidelity. The Table~\ref{tab:meas} data also show the importance of connectivity as omitting one of the CNOTs causes the algorithmic error to increase appreciably.

One of the main questions about 3Q RB is how much new information does it convey, i.e., can 3Q errors be predicted from the 1Q and 2Q errors (more specifically, the 1Q and 2Q depolarizing rates) since the 3Q Clifford gates are built from the set of one and two-qubit gates? To answer this question we calculate the predicted 3Q decay parameter $\alpha$ (converting to EPC using Eqn.~\ref{eqn:epc}),
\begin{eqnarray}
	\alpha_{3Q} & = & \frac{\alpha_{1Q}^{N_1/3} \alpha_{2Q}^{2N_3/3}}{7} \left(1+3\alpha_{1Q}^{N_1/3}\alpha_{2Q}^{N_2/3}+ \nonumber \right. \\
		    & & \left. 3\alpha_{1Q}^{2N_1/3}\alpha_{2Q}^{N_2/3} \right)  \label{eqn:3qepc}
\end{eqnarray}
where $N_2$ ($N_1$) is the number of 2Q (1Q) gates per 3Q Clifford, and $p_1=1-\alpha_1 (p_2=1-\alpha_2)$ are the 1Q (2Q) depolarizing probabilities. For simplicity we assume that all 1Q gates and 2Q gates have the same depolarizing probability; see~\cite{mckay:2018} for the general form of Eqn.~\ref{eqn:3qepc} and details of the derivation. The values discussed previously for $N_1$ and $N_2$ did not consider the finite duration of gates. In reality, there will be idle periods on some qubits and characterizing idle periods as one-qubit gates, $N_1=34.7$ ($N_1=67.9$) for all-to-all (limited) connectivity. This is the number used for predicting the 3Q EPC.

For the 1Q and 2Q depolarizing probabilities in Eqn~(\ref{eqn:3qepc}) we use two sets of numbers from Table~\ref{tab:meas} and compare the predicted to measured 3Q EPC as shown in Table~\ref{tab:pred}. The first set are the coherence limited EPGs. Unsurprisingly, the measured 3Q EPC is much higher than the coherence limited error, indicating the the majority of errors are due to unwanted and uncompensated terms in the Hamiltonian such as crosstalk. The second set of numbers is from 2Q-1Q simultaneous RB, which should be the most accurate measure of primitive gate errors. Indeed, for calibration A the estimate of the 3Q EPC from 2Q-1Q RB is accurate for both all-to-all and limited connectivity. Howevever, in the case of calibration B, there is very little agreement between the predicted and measured 3Q EPC, demonstrating the utility of 3Q RB as a unique measurement of multiqubit fidelity sensitive to subtle errors that are not fully revealed by benchmarking the primitive gates.

In the system studied here, the dominant crosstalk error is due to unwanted ZZ interactions. By calibrating the average ZZ shift into the qubit frequencies (calibration A) this error is mitigated to the point that the remaining error is dominated by stochastic terms that equally affect 2Q-1Q simultaneous and 3Q RB. However, when the ZZ shifts are not compensated (calibration B), their effect depends on the structure of the RB sequence. The errors measured from 2Q-1Q RB are lower because there is an aspect of dynamical decoupling that works to cancel the ZZ terms. For example, during 1Q RB the qubit changes directions on the Bloch sphere every approximately 50~ns independent of the 2Q Clifford gates applied to the other two qubits. However, the full 3Q Clifford gates have idle periods on the spectator qubits while the other qubits perform the 2Q gate (this is schematically illustrated in Fig.~\ref{fig:1} d.). The structure of the 3Q Clifford gate is not unique and certain constructions may amplify or attenuate different error terms; investigating such constructions in detail is left for future study.

In conclusion, we demonstrate, for the first time, 3Q RB and subset 2Q-1Q simultaneous RB. Although there is no true primitive three-qubit gate, 3Q RB measures a fidelity that is not captured by the one- and two-qubit gate metrics. As systems continue to increase in size and crosstalk terms dominate error, metrics such as 3Q RB will play an important role in benchmarking the true algorithmic fidelity of these large systems.

\begin{acknowledgments}
We thank Firat Solgun, Markus Brink, Sami Rosenblatt and George Keefe for modeling and fabricating the device. We thank Lev Bishop, Andrew Cross, Easwar Magesan and Antonio Corcoles for discussions and manuscript comments. We thank Christopher Wood and Sergey Bravyi for help generating the Clifford gates. This work was supported by the Army Research Office under contract W911NF-14-1-0124.
\end{acknowledgments}

\bibliography{threeq}

\end{document}